\documentstyle[aps,twocolumn,epsfig]{revtex}

\begin{document}

\draft

\title{Landau damping in dilute Bose gases}

\author{ L. P. Pitaevskii$^{1,2,3}$ and S. Stringari$^{3}$}

\address{$^{1}$Department of Physics, Technion, 32000 Haifa, Israel}
\address{$^{2}$Kapitza Institute for Physical Problems, 117454 Moscow,
Russia}
\address{$^{3}$Dipartimento di Fisica, Universit\`a di Trento, \protect\\
and Istituto Nazionale di Fisica della Materia, I-38050 Povo, Italy}
\date{August 7, 1997}

\maketitle

\begin{abstract}

{\it Landau damping in weakly interacting Bose gases is investigated by
means of perturbation theory. Our approach  points out the crucial role
played by Bose-Einstein condensation and yields an explicit expression for
the decay rate of elementary excitations in both uniform and non uniform
gases. Systematic results are derived for the   phonon width in homogeneous
gases interacting with repulsive forces. Special attention is given to the 
low and high temperature regimes.}

\end{abstract}


\narrowtext

The collective excitations of gases of alkali atoms confined in magnetic
traps  have been the object of recent experimental measurements \cite{exp}.
The observed frequencies very well agree with the theoretical predictions
of mean field theory \cite{theory}, confirming the validity of the
Bogoliubov scheme, later extended by Gross and Pitaevskii to the
non homogeneous case, for describing the dynamic behaviour of dilute Bose
gases. Conversely, the damping mechanisms associated with such excitations
are not well understood and still represent a challenging question for
theoretical investigation.

The damping of collective modes can have various origins. At
$T=0$ it  can arise because of decay  into two or more excitations with
lower energy. This mechanism, which is well understood
in uniform Bose superfluids \cite{beliaev}, is not active for the lowest
modes of a trapped gas, because of  discretization of levels. Damping can
be due to other non linear phenomena of classical or quantum nature
\cite{kagan}. Experiments with magnetic traps however point out the
occurrence of damping  in an almost  linear regime where such effects
should be negligible. At finite temperature damping can be due to
collisional effects of dissipative type. These effects are expected to be
important at high temperature and density. Another mechanism, holding in
the collisionless regime, is Landau damping which occurs when the
collective excitations can decay due to coupling with  transitions
associated with other elementary excitations and occurring  at the same
frequency. Landau damping is not associated with thermalization processes 
and can be well described in the framework of mean field theory (see, for
example, \cite{LP} \S \S 28-29, 34 and \cite{stefano}). In a Bose gas it
arises only at finite temperature. Its possible relevance  to
explain the experimental data in  trapped Bose gases has been recently
proposed by Liu and Schieve \cite{dam}.

The purpose of this letter is to develop the microscopic formalism of
Landau damping in a dilute Bose gas. The aim is twofold. On the one hand we
provide a systematic description of Landau damping in homogeneous systems
covering the regimes of low temperature, first investigated  by Hohenberg
and Martin \cite{HM}, and the one at higher temperatures first investigated
by Szepfalusy and Kondor \cite{SK}. On the other hand we provide a
formalism suitable for the calculation of Landau damping in trapped  gases 
where the classification of elementary excitations  significantly differs
from the one of uniform systems. Our formalism is based on
perturbation theory and points out the crucial role played by Bose-Einstein
condensation. Previous approaches were based on kinetic theory for
superfluids \cite{LP} or on the use of Green's function techniques
\cite{HM}.

Let $E$ be the energy of the system associated with the occurrence of a
classical oscillation of frequency $\omega$ induced, for example, by some
external drive as happens in the experiments of ref.\cite{exp}. By
classical oscillation we mean that the number of  quanta of oscillation
is very large. Due to  interaction effects, the thermal component of the
gas can either absorb or emit quanta of this oscillation, thereby giving
rise to the following expression for the energy loss
\begin{equation}
\dot E _{os} = -\hbar \omega (W^{(a)}-W^{(e)}) \; .
\label{1}
\end{equation}
In the above equation $W^{(a)}$ and $W^{(e)}$ are, respectively, the
probabilities of absorption and emission of the corresponding quantum
$\hbar \omega$ given, in perturbation theory, by
\begin{equation}
W = \pi \sum_{i,k} \mid \langle k \mid V_{int}\mid  i \rangle
\mid^2 \; ,
\label{2}
\end{equation}
where the matrix element is associated with a transition in which the
$i$-th excitation, available in the system because of thermal activation,
is transformed into the $k$-th one and $E_k = E_i +\hbar \omega$ in the
case of absorption (a), and   $E_k = E_i -\hbar \omega$ in the case of
emission (e).

In second quantization the interaction term can be written in
the following way
\begin{equation}
V _{int}= \frac{g}{2} \int \! d{\bf r} \ \psi ^{\dagger } \psi
^{\dagger }\psi \psi \; ,
\label{Vint}
 \end{equation}
where the coupling constant $g$ is related to the s-wave scattering length
through the usual formula $g=4\pi \hbar^2a / m$. The field operator $\psi$
can be written as $\psi =\psi _0+\delta \psi$ where $\psi_0$ is the order
parameter at equilibrium while $\delta \psi$ characterizes its fluctuations
which can be expressed in terms of the annihilation ($\alpha$) and creation
($\alpha^{\dagger}$) operators relative to the elementary modes of the
system in the usual form
\begin{equation}
\delta \psi = \sum _{j} \left ( u_j(\bbox{r}) \alpha _j
 + v_j^*(\bbox{r})\alpha _j ^{\dagger }\right )
 \label{deltapsi}
 \end{equation}
and $u,v$ can be determined, for example, using Bogoliubov theory or its
extensions to finite temperature \cite{griffin,giorgini}. In the sum
(\ref{deltapsi}) one should distinguish between the collective excitation
whose decay is the object of the present work and for which we use the
notation $u_{osc}, v_{osc}, \alpha_{osc}, \alpha_{osc}^{\dagger}$, and the
other elementary modes which are thermally excited and for which we will
use the indices $i,k$ already employed in  (\ref{2}). By keeping in
(\ref{Vint}) only terms linear in $\alpha_{osc},(\alpha_{osc}^{\dagger})$ 
and in the product $\alpha_k^{\dagger}\alpha_i$
($\alpha_k\alpha_i^{\dagger}$), we investigate the desired process where a
quantum of oscillation $\hbar \omega$ is annihilated (created) and the
$i$-th excitation is transformed into the $k$-th one. The same formalism
permits to investigate also other decay processes where both the
excitations $i$ and $k$ are  created (Beliaev decay of a phonon
into two phonons \cite{beliaev}) and $E_k+E_i = \hbar \omega$. In our case
we find the following result for the rate of energy loss:
\begin{equation}
\dot E =-\omega 2\pi \sum _{ik} \mid A_{ki} \mid ^2 \delta
(E_k-E_i-\hbar \omega ) (f_k-f_i) \; .
\label{dot}
\end{equation}
 where $E=\hbar\omega n_{osc}$ is the energy of the classical oscillation
($n_{osc} \gg 1$) and
\begin{eqnarray}
A_{ki} = 2 g \int \! d{\bf r}  \ & \psi _0 &
[(u_k^*v_i+v_k^*v_i+u_k^*u_i) u_{osc} \nonumber \\
& + &(v_k^*u_i+ v_k^*v_i+u_k^*u_i)v_{osc} ]
\label{A}
\end{eqnarray}
is the relevant matrix element for the process. In deriving (\ref{dot}) we
have assumed that at equilibrium the states $i,k$ are thermally occupied
with the usual Bose factor $f_j=[\exp(\beta E_j)-1]^{-1}$ .
Introducing the damping rate through the relation $\dot E = -2\gamma E$ we
finally obtain the relevant formula
\begin{equation}
\gamma=-\omega \pi \sum _{ik} \mid A_{ki} \mid ^2 \delta
(E_k-E_i-\hbar \omega ) \ \frac{\partial f(E_i)}{\partial E} \; ,
\label{gamma2}
\end{equation}
where we have further assumed $\hbar\omega \ll T$.
Eqs. (\ref{A},\ref{gamma2}) can be further
simplified by calculating the relevant matrix elements in semiclassical
approximation and can be used to calculate the damping
rate of collective excitations in a trapped Bose gas. This calculation will
be the object of a future work.  In the following we will use them for a
systematic discussion of Landau damping in a homogeneous gas where all the
ingredients take a simplified form. In this case $\psi_0 = \sqrt{n_0}$ is 
constant and $u,v$ are plane
wave functions:
$u= U({\bf p}) \exp(i\bbox{p \cdot r})/\sqrt{V}$ and
$v= V({\bf p}) \exp(i\bbox{p \cdot r})/ \sqrt{V}$.
After some straightforward algebra,  we can rewrite the rate of the
collective excitation propagating with momentum ${\bf q}$ in the following
way:
\begin{equation}
\gamma=-\omega \pi \int \frac{d{\bf p}_i}{(2\pi )^3} 
\ | B_{ki} |^2 \delta (E_k-E_i- \omega )
\frac{\partial f(E_i)}{\partial E} \; ,
\label{gamma5}
\end{equation}
where
\begin{eqnarray}
B_{ki} \  =  \ & 2 & g \sqrt{n_0} \ \{ [ U(E_k) V(E_i) +V(E_k)V(E_i)
\nonumber \\
               & + & U(E_k)U(E_i)] U_{osc} + [ V(E_k)V(E_i)
\nonumber \\
               & + & V(E_k)V(E_i)+U(E_k)U(E_i) ] V_{osc} \} \; ,
 \label{B}
 \end{eqnarray}
\begin{eqnarray}
&\;& U^2(E) = 1+V^2(E)=
\frac
{\left (E^2+g^2n_0^2) \right )^{1/2}
+ E}{2E}
\label{UV} \\
&\;& U(E)V(E) = - \frac{gn_0}
{2E} \;\; \nonumber
 \end{eqnarray}
and
\begin{equation}
E({\bf p}) = \left( \left( \frac{p^2}{2m}
+gn_0 \right)^2 - g^2n_0^2 \right)
^{1/2} \; .
 \label{E2}
 \end{equation}
is the  dispersion law of the elementary excitations in the so
called Popov approximation \cite{griffin,giorgini}, depending on the
corresponding value of the momentum ${\bf p}$. In these equations $n_0
\equiv n_0(T)$ is the condensate density at  temperature $T$ and the
momenta {\bf p} of the $i$-th and $k$-th excitations  satisfy
the condition ${\bf p}_k={\bf p}_i + {\bf q}$,  following from momentum
conservation.

A further simplification is obtained if one considers the damping for the 
low frequency  excitations, i. e. for phonons
(${\bf q} \to 0$). In this case, one has
\begin{equation}
U_{osc}(\omega ) \approx \left ( \frac{gn_0}{2\omega }\right )^{1/2}+
\frac{1}{2}\left ( \frac{\omega }{2gn_0}\right )^{1/2}
\label{U}
\end{equation}
and
\begin{equation}
V_{osc}(\omega ) \approx -\left ( \frac{gn_0}{2\omega }\right )^{1/2}+
\frac{1}{2}\left ( \frac{\omega }{2gn_0}\right )^{1/2} \; .
\label{V}
\end{equation}

In the same limit ${\bf q} \to 0$, using momentum
conservation, one can write  $E_k-E_i \approx v_g\omega \cos(\theta )/c$, 
where $\theta $ is the angle between $\bbox{p}_i$ and $\bbox{q} $,   $v_g $
is the group velocity of
the $i$-th excitation and  $c$ is  the velocity of sound.
After integration with respect to $\theta $ and
some straightforward algebra one finds the useful result
\begin{equation}
\gamma=-c \omega \int \frac{p^2 dp}{4\pi v_g} 
\mid C \mid ^2 \frac{\partial  f(E)}{\partial E}
\; ,
\label{gamma7}
\end{equation}
where
\begin{equation}
C(E)=\sqrt{2g} [  U^2(E)+ V^2(E)+U(E)V(E)  + E
\frac{\partial U^2(E) }{\partial E} ] \; .
\label{C}
\end{equation}

Result (\ref{gamma7}) permits to explore explicitly the $T$-dependence of
Landau damping. At low $T$ ($T \ll \mu $, where $\mu
=gn_0(0) $ is the $T=0$ value of the chemical potential),
 one finds  the result
\begin{equation}
\gamma= \frac{27\pi}{16} \frac{\omega \rho _n}{\rho} \; ,
\label{gamma8}
\end{equation}
where $\rho _n$ is the normal part density of the phonon
gas \cite{9}, \S 23:
\begin{equation}
\rho _n = \int \frac{p^4dp}{6\pi^2}
\left( - \frac{\partial f(E)}{\partial E}\right) \ = \
\frac{2\pi ^2T^4}{45\hbar ^3c^5} \; .
\label{rn}
\end{equation}
Eq. (\ref{gamma8}) has been first derived by P.Hohenberg and P.Martin
\cite{HM}. This equation can be also obtained following the  method of
\cite{AK}. (See \cite{LP}, Problem to \S 77.)

A second important regime occurs at relatively high $T$ ($T \gg \mu$).
Usually at such temperatures the thermodynamic
quantities are determined by the excitations with $E \sim T$. In the
integral (\ref{gamma7}) however the relevant excitations turn out to have
energies $E \sim \mu \sim gn_0 \ll T$, so that  one can use the
''Rayleigh-Jeans''  limit of the distribution function,
$f(E) \approx T/E$. As a consequence  the resulting
dependence of $\gamma $ is linear in $T$ and  integration of (\ref{gamma7})
gives the result
 \begin{equation}
\gamma = \frac{3\pi }{8}\frac{Taq}{\hbar^2}
\label{high}
\end{equation}
where we have used the expression $c^2 = gn_0/m $ for the sound velocity.
This regime has been previously investigated by Szepfalusy and Kondor
\cite{SK} and more recently by Hua Shi \cite{H}. Our result
coincides with the one obtained in \cite{H} but the numerical coefficient
slightly differs from the one of \cite{SK}. It is worth noting that result
(\ref{high}) does not depend on the value of the condensate fraction $n_0$.
The above "high" T regime has been recently employed in \cite{dam} to
provide a quantitative estimate of the width of
the collective excitations of a dilute Bose gas trapped by a  harmonic
potential. Estimating  the momentum $q$ of the excitation as
$\hbar\omega/c$ where $\omega$ is the frequency of the excitation and
$c= ( \hbar / m) [ 4\pi an_0(r=0)]^{1/2}$ is the
value of the sound velocity calculated in the center of the trap, the
authors of \cite{dam} have obtained values for $\gamma$ in
semi-quantitative agreement with the experimental findings \cite{exp}. This
agreement suggests that  Landau damping could represent an
important mechanism to explain the decay of elementary excitations of
trapped Bose gases. Notice however that in a non uniform system the
elementary excitations cannot be  described in terms of plane wave
functions and that Landau damping should be consequently calculated
starting directly from the general equation (7), which involves matrix
elements and eigenvalues relative to the eigenstates of the system. The
presence of the trapping potential can influence the damping
mechanism in a deep way. For example the dipole oscillation does not
exhibit any damping in the presence of harmonic trapping.

 It is finally interesting to discuss the validity of the ``high"
temperature expansion  (\ref{high}). To this purpose it is
useful to write the ratio $\gamma/\omega$ in the form (see Eq. (14))
 \begin{equation}
\frac{\gamma}{\omega} = (a^3n_0(T))^{1/2}F(\tau)
\label{fraction}
\end{equation}
where $\tau = T/mc^2(T)$ is a dimensionless variable and the function $F$
is given by
 \begin{equation}
F(\tau) = { 4 \sqrt{\pi} \over \tau} \int_0^{\infty} \!  dx \
(1-\frac{1}{2u}-\frac{1}{2u^2})^2 (e^{x/2\tau}-e^{-x/2\tau})^{-2}
\label{F} \end{equation}
with $u(x)=(1+x^2)^{1/2}$. For large $\tau$ (high temperatures) the
function $F$ takes the asymptotic value $F\to \frac{3}{4}\pi^{3/2}\tau$,
consistent with result (\ref{high}). In Fig. 1 we plot the function $F$
versus $\tau$. One can see that $F$ approaches the asymptotic linear law
rather slowly. For example for temperatures    $kT \sim mc^2$
($\tau \sim 1$) it differs from it by a factor 2. This suggests that the
linear approximation (\ref{high}) should be employed with care. In fact in
the available traps \cite{exp} the value of $mc^2$, with the sound velocity
calculated in the center of the trap, corresponds  to about $  0.4
T_c$ \cite{giorgini}, where $T_c$ is the critical temperature for
Bose-Einstein condensation in the presence of the trapping potential, and
consequently the condition  $\tau \gg 1$ is never reached
below $T_c$.  This should be taken into account for quantitative estimates
of the width using the present formalism.

It is a pleasure to thank S. Giorgini for many useful discussions.

After completing the present work we have been informed of a similar
analysis of Landau damping \cite{ciocco} carried out in dilute Bose gases
using semiclassical theory.

\begin{figure}
\caption{Function $F$ plotted as a function of $\tau$  (full line). The
linear behavior $F\to (3/4) \pi^{3/2}\tau$ is also reported (dashed line).}
\end{figure}

\end{document}